\documentclass[10pt,twocolumn]{IEEEtran}

\usepackage{cite}
\usepackage{graphicx}
\usepackage{epsfig}
\usepackage{psfrag}
\usepackage{subfigure}
\usepackage{url}
\usepackage{stfloats}
\usepackage{amsmath}
\usepackage{color}
\usepackage{amssymb}
\usepackage{algorithm}
\usepackage{algorithmic}
\usepackage{multirow}
\usepackage{array}
\usepackage{booktabs}
\usepackage{multirow}

\begin{document}
%
\title{Spatiotemporal Sparse Bayesian Learning with Applications to Compressed Sensing of Multichannel Physiological Signals}

\author{Zhilin Zhang$^*$, \IEEEmembership{Member, IEEE}, Tzyy-Ping Jung, \IEEEmembership{Senior Member, IEEE}, Scott Makeig, \IEEEmembership{Member, IEEE}  \\Zhouyue Pi, \IEEEmembership{Senior Member, IEEE}, Bhaskar D. Rao, \IEEEmembership{Fellow, IEEE}
\thanks{Manuscript received May 7, 2013; revised February 18, 2014; accepted April 13, 2014. The work was supported by NSF grants CCF-0830612, CCF-1144258 and DGE-0333451, and was in part supported by Army Research Lab, Army Research Office, Office of Naval Research and DARPA. \emph{Asterisk indicates corresponding author}.}
\thanks{Z. Zhang was with the Department of Electrical and Computer Engineering, University of California at San Diego, La Jolla, CA 92093-0407, USA. He is now with the Emerging Technology Lab, Samsung Research America -- Dallas, 1301 East Lookout Drive, Richardson, TX 75082, USA. Email: zhilinzhang@ieee.org}
\thanks{T.-P. Jung and S. Makeig are with the Swartz Center for Computational
Neuroscience and the Center for Advanced Neurological Engineering, University of California at San Diego, La Jolla, CA 92093-0559, USA.}
\thanks{Z. Pi is with the Emerging Technology Lab, Samsung Research America -- Dallas, 1301 East Lookout Drive, Richardson, TX 75082, USA.}
\thanks{B.D. Rao is with the Department of Electrical and Computer Engineering, University of California at San Diego, La Jolla, CA 92093-0407, USA.}
}

\markboth{Published in IEEE Transactions on Neural Systems and Rehabilitation Engineering, Vol. 22, No. 6, pp. 1186--1197, November 2014}{Zhang \MakeLowercase{\textit{et al.}}: }

\maketitle

\begin{abstract}

Energy consumption is an important issue in continuous wireless telemonitoring of physiological signals. Compressed sensing (CS) is a promising framework to address it, due to its energy-efficient data compression procedure. However, most CS algorithms have difficulty in data recovery due to non-sparsity characteristic of many physiological signals.  Block sparse Bayesian learning (BSBL) is an effective approach to recover such signals with satisfactory recovery quality. However, it is time-consuming in recovering multichannel signals, since its computational load almost linearly increases with the number of channels.

This work proposes a spatiotemporal sparse Bayesian learning algorithm to recover multichannel signals simultaneously. It not only exploits temporal correlation within each channel signal, but also exploits inter-channel correlation among different channel signals. Furthermore, its computational load is not significantly affected by the number of channels. The proposed algorithm was applied to brain computer interface (BCI) and EEG-based driver's drowsiness estimation. Results showed that the algorithm had both better recovery performance and much higher speed than BSBL. Particularly, the proposed algorithm  ensured that the BCI classification and the drowsiness estimation had little degradation even when data were compressed by 80\%, making it very suitable for continuous wireless telemonitoring of multichannel signals.

\end{abstract}


%

\IEEEpeerreviewmaketitle

\section{Introduction}
\label{sec:intro}

Compressed sensing (CS) \cite{Candes2008CS} has been drawing increasing attention in the wireless telemonitoring of physiological signals as an emerging data compression methodology \footnote{The CS technique can be used for data compression and signal sampling \cite{Candes2008CS}. In this paper the use of CS for data compression/de-compression is considered. But note the proposed algorithm can be also used as a signal recovery method in CS-based sampling. } \cite{Aviyente2007EEG,Pinheiro2010,Dixon2012,Chen2012design,Zhang2012TBME,Zhang2012Letter}. It has been shown that CS, compared to traditional data compression methodologies, consumes much less energy and power \cite{Mamaghanian2011}, saves lots of on-chip computational resources \cite{LiuZhang2013}, and is robust to packet loss during wireless transmission \cite{fauvel2014energy}. Thus it is attractive to wireless body-area networks for ambulatory monitoring.

\subsection{CS Models}

The basic CS framework \cite{Candes2008CS}, also called the single measurement vector (SMV) model, can be expressed as
\begin{eqnarray}
\mathbf{y}= \mathbf{\Phi} \mathbf{x} + \mathbf{v},
\label{equ:model_CSbasic}
\end{eqnarray}
where, in the context of data compression, $\mathbf{x} \in \mathbb{R}^{M \times 1}$ is a single-channel signal, $\mathbf{\Phi} \in \mathbb{R}^{N \times M} (N < M)$ is a user-designed measurement matrix, $\mathbf{v} \in \mathbb{R}^{N \times 1}$ is sensor noise, and $\mathbf{y} \in \mathbb{R}^{N \times 1}$ is the compressed signal. This compression task is performed in sensors of a wireless body-area network. Then, the compressed signal $\mathbf{y}$, through Bluetooth and Internet, is sent to a remote terminal. At the terminal, the original signal is recovered by a CS algorithm using the shared measurement matrix $\mathbf{\Phi}$, namely \footnote{There are other mathematical expressions, which are equivalent given suitable values for regularization parameters.},
\begin{eqnarray}
\widehat{\mathbf{x}} = \arg\min_{\mathbf{x}} \| \mathbf{y} - \mathbf{\Phi} \mathbf{x}\|_2^2  + \lambda g(\mathbf{x}),
\label{equ:solution_CSbasic}
\end{eqnarray}
where $\lambda$ is a regularization parameter, and $g(\mathbf{x})$ is a penalty function of $\mathbf{x}$. The most popular penalty may be the $\ell_1$-minimization based penalty, namely $g(\mathbf{x}) = \|\mathbf{x}\|_1$. This method is called \emph{signal recovery in the original domain}.

When the original signal $\mathbf{x}$ is sufficiently sparse (i.e., only a few entries of $\mathbf{x}$ are nonzero), many CS algorithm can exactly recover $\mathbf{x}$ from $\mathbf{y}$ in the absence of noise $\mathbf{v}$
or with high quality in the presence of noise \footnote{Admittedly, when $\mathbf{x}$ is sparse, it is trivial to use CS for data compression, because one can just send nonzero entries (and associated locations) of $\mathbf{x}$ to a remote terminal and then recover it over there. When $\mathbf{x}$ is non-sparse, directly using the recovery method (\ref{equ:solution_CSbasic}) results in failure for existing CS algorithms. Thus the recovery method (\ref{equ:solution_CSbasic}) is rarely used by CS algorithms. But the block sparse Bayesian learning can adopt this method (\ref{equ:solution_CSbasic}) to recover a non-sparse $\mathbf{x}$ with correlated entries (with very small errors)\cite{Zhang2012TBME}.}. If $\mathbf{x}$ is not sparse, one can seek a dictionary matrix $\mathbf{D}$ such that $\mathbf{x}$ can be sparsely represented under the dictionary matrix, i.e., $\mathbf{x} = \mathbf{Dz}$, where the representation coefficients $\mathbf{z}$ are sparse. The dictionary matrix $\mathbf{D} \in \mathbb{R}^{M \times m} (M \leq m)$ can be formed from orthonormal bases of known transforms such as discrete wavelet transform or discrete Cosine transform (DCT), or can be learned from data using dictionary learning \cite{kreutz2003dictionary}. Then a CS algorithm  recovers the original signal according to:
\begin{eqnarray}
\widehat{\mathbf{x}} = \mathbf{D} \widehat{\mathbf{z}} \quad \text{with} \quad \widehat{\mathbf{z}} = \arg\min_{\mathbf{z}} \| \mathbf{y} - \mathbf{\Omega} \mathbf{z}\|_2^2  + \lambda g(\mathbf{z})
\label{equ:model_CSbasic_withD}
\end{eqnarray}
where $\mathbf{\Omega} \triangleq \mathbf{\Phi} \mathbf{D}$. The method is called \emph{signal recovery in a transformed domain}.

The basic CS framework has been widely studied for data compression/decompression of biosignals \cite{Aviyente2007EEG,Abdulghani2012EEG,Chen2012design,Zhang2012TBME,Zhang2012Letter,mohsina2013gabor,pant2013compressive}. For example, Aviyente \cite{Aviyente2007EEG} studied the use of Gabor dictionary matrix for EEG. Later Abdulghani et al. \cite{Abdulghani2012EEG} further investigated various kinds of dictionary matrices. Instead of using the popular $\ell_1$-minimization based penalty, other more effective penalties were proposed, such as the block-sparsity with intra-block correlation \cite{Zhang2012Letter,Zhang2012TSP}, the analysis prior formulation \cite{mohsina2013gabor}, and the sparsity on second-order difference \cite{pant2013compressive}. Chen et al. \cite{Chen2012design} proposed an energy-efficient digital implementation of CS architecture for data compression in wireless sensors. Using a field programmable gate array (FPGA) platform, Liu et al. \cite{LiuZhang2013} showed that CS, when compared to a low-power wavelet compression procedure, can largely save energy, power, and other on-chip computational resources.

In addition to the SMV model (\ref{equ:model_CSbasic}), another widely studied CS model is the multiple measurement vector (MMV) model \cite{Cotter2005}, an extension of the SMV model. It can be expressed as follows:
\begin{eqnarray}
\mathbf{Y}= \mathbf{\Phi} \mathbf{X} + \mathbf{V},
\label{equ:model_MMVbasic}
\end{eqnarray}
where $\mathbf{Y} \in \mathbb{R}^{N \times L}$, $\mathbf{X} \in \mathbb{R}^{M \times L}$ and $\mathbf{V} \in \mathbb{R}^{N \times L}$ are matrices. A key assumption in the MMV model is that $\mathbf{X}$ is row sparse, namely only a few rows of $\mathbf{X}$ are nonzero rows. Similar to (\ref{equ:solution_CSbasic}) and (\ref{equ:model_CSbasic_withD}), the estimate of $\mathbf{X}$ is given by
\begin{eqnarray}
\widehat{\mathbf{X}} = \arg\min_{\mathbf{X}} \| \mathbf{Y} - \mathbf{\Phi} \mathbf{X}\|_2^2  + \lambda f(\mathbf{X}),
\label{equ:solution_MMV}
\end{eqnarray}
or given by
\begin{eqnarray}
\widehat{\mathbf{X}} = \mathbf{D} \widehat{\mathbf{Z}} \quad \text{with} \quad \widehat{\mathbf{Z}} = \arg\min_{\mathbf{Z}} \| \mathbf{Y} - \mathbf{\Omega} \mathbf{Z}\|_2^2  + \lambda f(\mathbf{Z})
\label{equ:solution_MMV_withD}
\end{eqnarray}
where $\mathbf{\Omega} \triangleq \mathbf{\Phi} \mathbf{D}$, and $\mathbf{D}$ is a dictionary matrix. $f(\mathbf{X})$ is a penalty encouraging row-sparsity of $\mathbf{X}$. One popular penalty is the $\ell_1/\ell_2$-minimization based penalty, namely $f(\mathbf{X})=\sum_{i=1}^M \| \mathbf{X}_{i \cdot} \|_2$. In (\ref{equ:solution_MMV_withD}) it is assumed that $\mathbf{Z}$ is row-sparse.

Compared to recovering $\mathbf{X}$ column by column, i.e., treating (\ref{equ:solution_MMV}) [or (\ref{equ:solution_MMV_withD})] as $L$ individual sub-problems, the joint recovery as in (\ref{equ:solution_MMV}) [or (\ref{equ:solution_MMV_withD})] can greatly improve the recovery quality of  $\mathbf{X}$ \cite{Cotter2005,Eldar2010average}. Aviyente \cite{Aviyente2007EEG} explored this model to jointly recover multichannel EEG signals. Polania et al. \cite{polania2011compressed} explored this model to jointly recover multichannel ECG signals. However, the benefit of the MMV model is largely compromised if  columns of $\mathbf{X}$ exhibit inter-vector correlation; the benefit even almost disappears when the inter-vector correlation is very high \cite{Zhang2011IEEE}.

Recently, by proposing the T-MSBL algorithm \cite{Zhang2011IEEE}, we showed that suitably exploiting the inter-vector correlation can greatly alleviate its negative effect. Particularly, in noiseless environments, under mild conditions the negative effect disappears no matter how large the inter-vector correlation is (as long as the correlation is not $\pm$ 1).  This algorithm motivated the development of the spatiotemporal algorithm presented in this paper.

\subsection{Challenges in the Use of CS for Wireless Telemonitoring}

It is worth pointing out that most CS algorithms may not be used for energy-efficient wireless telemonitoring especially ambulatory monitoring, due to several challenges \cite{Zhang2013Asilomar,milenkovic2006wireless,Martin2000issues}.

One challenge comes from the strict energy constraint. A wireless telemonitoring system is generally battery-operated. This situation with other constraints (e.g.,  wearability and device cost) requires that the compression procedure should be as simple as possible. In other words, the pre-processing such as filtering, peak detection, and dynamical thresholding, is not favored, since they increase circuitry complexity and cost extra energy. In fact, the data compression stage should be very simple. Lots of evidence have shown that the energy-saving advantage of CS over  conventional data compression methods might be true only when the measurement matrix $\mathbf{\Phi}$ was a \emph{sparse binary matrix}; when $\mathbf{\Phi}$ was a random Gaussian matrix or other kinds of matrices, the advantage disappeared.

Another challenge comes from strong artifacts caused by human movement during data recording. The goal of wireless telemonitoring is to allow people to move freely. Thus, the collected physiological signals are inevitably contaminated by strong artifacts caused by muscle movement and electrode motion. As a result, even a sparse signal can become non-sparse in the time domain and also non-sparse in  transformed domains \cite{Zhang2013Asilomar}. The non-sparsity seriously degrades CS algorithms' performance, resulting in their failure \cite{Zhang2012TBME}. Therefore, CS algorithms generally need to remove artifacts before compression. But this greatly increases circuitry complexity, and conflicts with the energy constraint. The conflict is more sharp in some scenarios such as ambulatory telemonitoring.

Very recently, we proposed using the block sparse Bayesian learning (BSBL) framework \cite{Zhang2012TSP} for CS of non-sparse physiological signals, and achieved success in  telemonitoring of fetal ECG \cite{Zhang2012TBME} and single-channel EEG \cite{Zhang2012Letter}. The significant innovation in those works is that, instead of using the mentioned pre-processing methods or seeking optimal dictionary matrices, we proposed a completely different approach: namely recovering  non-sparse signals directly without resorting to optimal dictionary matrices or pre-processing methods. The key element in BSBL is  exploitation of correlation structures of a signal.

However, BSBL is designed for recovering single-channel signals. When recovering multichannel signals, BSBL has to recover the signals channel by channel, which is time-consuming and thus not suitable for real-time telemonitoring of multichannel signals. Besides, for many multichannel physiological signals, there is strong correlation among signals of different channels. Exploiting the inter-channel correlation is necessary and very beneficial. Unfortunately, BSBL ignores it.

\subsection{Summary of the Work}

The work introduces a spatiotemporal sparse model to the field of CS. This model is an extension of the classic multivariate Bayesian variable selection model \cite{Brown1998multivariate}, and was recently used in overdetermined multivariate regression models to identify predictors by exploiting nonlinear relationships between predictors and responses \cite{Wan2014identifying}. However, this model has not been studied in CS.

Based on this model, we derive an expectation-maximization based spatiotemporal sparse Bayesian learning algorithm, and apply it to CS of multichannel signals. This algorithm has several advantages.
      \begin{itemize}
        \item It can efficiently exploit  temporal correlation of each channel signal and inter-channel correlation among different  channel signals to improve recovery performance. As we will see later, exploiting the inter-channel correlation is very important in CS of multichannel signals.

        \item It has the ability to recover non-sparse correlated signals, and signals with less-sparse representation coefficients,  a desired ability for wireless telemonitoring of physiological signals.

        \item Compared to BSBL, it not only has  better recovery performance, but also has higher speed. Its computational load is not significantly affected by the number of channels, an obvious advantage over BSBL. Thus it is very suitable for CS of multichannel signals.

        \item Different from most CS algorithms, which require preprocessing before compressing raw data, the proposed algorithm does not require any preprocessing. Its compression procedure can be implemented by very simple circuits, thus costing ultra-low energy consumption. This is highly desired for long-term wireless telemonitoring of physiological signals.
      \end{itemize}

In experiments on steady-state visual evoked potential (SSVEP) based BCI and EEG-based driver's drowsiness estimation, the proposed algorithm ensured that the BCI classification and the drowsiness estimation on recovered data were almost the same as on original data, even when the original data were compressed by more than 80\%.

Some preliminary results were published in \cite{Zhang2013Asilomar}. The Matlab code of the proposed algorithm can be downloaded at \url{https://sites.google.com/site/researchbyzhang/stsbl}.

\subsection{Organization and Notations}

The rest of the paper is organized as follows. Section \ref{sec:model} describes the spatiotemporal sparse model. Section \ref{sec:algorithm} derives a spatiotemporal sparse Bayesian learning algorithm using the expectation-maximization method. Section \ref{sec:apply2CS} discusses some specific settings when applying the algorithm for CS of multichannel physiological signals. Section \ref{sec:experiments} presents experimental results on BCI and EEG-based driver's drowsiness estimation. Discussion and conclusion are given in the last two sections.

We introduce the notations used in this paper:
\begin{itemize}
  \item Bold symbols are reserved for vectors and matrices. Particularly, $\mathbf{I}_L$ denotes the identity matrix with size $L \times L$. When the dimension is evident from the context, for simplicity, we just use $\mathbf{I}$;
  \item $\| \mathbf{x} \|_1,\| \mathbf{x} \|_2,\| \mathbf{A} \|_\mathcal{F}$ denote  the $\ell_1$ norm of the vector $\mathbf{x}$, the $\ell_2$ norm of $\mathbf{x}$, and the Frobenius norm of the matrix $\mathbf{A}$, respectively;
  \item $\mathrm{diag}\{a_1,\cdots,a_M \}$ denotes a diagonal matrix with principal diagonal elements being $a_1,\cdots,a_M$ in turn; if $\mathbf{A}_1,\cdots,\mathbf{A}_M$ are  square matrices, then $\mathrm{diag}\{\mathbf{A}_1,\cdots,\mathbf{A}_M \}$ denotes a block diagonal matrix with principal diagonal blocks being $\mathbf{A}_1,\cdots,\mathbf{A}_M$ in turn;
  \item $\mathbf{A} \otimes \mathbf{B}$ represents the Kronecker product of the two matrices $\mathbf{A}$ and $\mathbf{B}$. $\mathrm{vec}(\mathbf{A})$ denotes the vectorization of the matrix $\mathbf{A}$ formed by stacking its columns into a single column vector. $\mathrm{Tr}(\mathbf{A})$ denotes the trace of $\mathbf{A}$. $\mathbf{A}^T$ denotes the transpose of $\mathbf{A}$;
  \item For a matrix $\mathbf{A}$, $\mathbf{A}_{i\cdot}$ denotes the $i$-th row, and $\mathbf{A}_{\cdot i}$ denotes the $i$-th column.  $\mathbf{A}_{[i],j}$ denotes the $i$-th block in the $j$-th column. $\mathbf{A}_{i,[j]}$ denotes the $j$-th block in the $i$-th row. When assuming all columns of $\mathbf{A}$ have the same block partition, $\mathbf{A}_{[i] \cdot}$ denotes the $i$-th block of all columns of $\mathbf{A}$.

\end{itemize}

\section{The Spatiotemporal Sparse Model}
\label{sec:model}

To enhance the readability of the paper, we first describe the spatiotemporal sparse model in this section, and delay the description of the proposed algorithm to the next section.

The spatiotemporal sparse  model is described as follows:
\begin{eqnarray}
\mathbf{Y}= \mathbf{\Phi} \mathbf{X} + \mathbf{V},
\label{equ:STmodel_original}
\end{eqnarray}
where $\mathbf{Y} \in \mathbb{R}^{N \times L}$, $\mathbf{\Phi} \in \mathbb{R}^{N \times M}$ \footnote{The model and the developed algorithm does not require $N < M$ or $N \geq M$. Thus they can be used for many other applications.}, and $\mathbf{X} \in \mathbb{R}^{M \times L}$. The matrices $\mathbf{Y}$ and $\mathbf{\Phi}$ are known. The goal is to estimate $\mathbf{X}$. In the context of data compression, the $l$-th column of $\mathbf{X}$, denoted by $\mathbf{X}_{\cdot l}$, is a segment of an original physiological signal in the $l$-th channel, and the $l$-th column of $\mathbf{Y}$ is the corresponding compressed segment.

The matrix $\mathbf{X}$ is assumed to have the following block structure:
\begin{eqnarray}
\mathbf{X}= \left[ \begin{array}{c}
 \mathbf{X}_{[1]\cdot}   \\
  \mathbf{X}_{[2]\cdot}  \\
     \vdots                \\
  \mathbf{X}_{[g]\cdot}    \end{array} \right]
\label{equ:STmodel_X_partition}
\end{eqnarray}
where $\mathbf{X}_{[i]\cdot} \in \mathbb{R}^{d_i \times L}$ is the $i$-th block of $\mathbf{X}$, and $\sum_{i=1}^g d_i = M$. For convenience, $\{d_1,\cdots,d_g\}$ is called the block partition. Among the $g$ blocks, only a few are nonzero.
The key assumption is that each block $\mathbf{X}_{[i]\cdot} (\forall i)$ is assumed to have spatiotemporal correlation. In other words, entries in the same column of $\mathbf{X}_{[i]\cdot}$ are correlated \footnote{In our data compression formulation, the correlation is a kind of temporal correlation of a channel signal.}, and entries in the same row of $\mathbf{X}_{[i]\cdot}$ are also correlated \footnote{In our data compression formulation, the correlation is called inter-channel correlation, and is also called spatial correlation.}.

The $i$-th block $\mathbf{X}_{[i]\cdot}$ is assumed to have the parameterized Gaussian distribution $p(\mathrm{vec}(\mathbf{X}_{[i]\cdot}^T); \gamma_i,\mathbf{B},\mathbf{A}_i)= \mathcal{N}(\mathbf{0},   (\gamma_i \mathbf{A}_i) \otimes \mathbf{B} )$. Here $\mathbf{B} \in \mathbb{R}^{L \times L}$ is an unknown positive definite matrix capturing the correlation structure in each row of $\mathbf{X}_{[i]\cdot}$. The matrix $\mathbf{A}_i \in \mathbb{R}^{d_i \times d_i}$ is an unknown positive definite matrix capturing the correlation structure in each column of $\mathbf{X}_{[i]\cdot}$. The unknown  parameter $\gamma_i$ is a nonnegative scalar, determining whether the $i$-th block is a zero block or not. Assuming the blocks $\{\mathbf{X}_{[i]\cdot}\}_{i=1}^g$ are mutually independent, the distribution of the matrix $\mathbf{X}$ is
\begin{eqnarray}
p(\mathrm{vec}(\mathbf{X}^T); \mathbf{B},\{\gamma_i,\mathbf{A}_i\}_i) =  \mathcal{N}(\mathbf{0},   \mathbf{\Pi}\otimes \mathbf{B})
\label{equ:x_pdf}
\end{eqnarray}
where $\mathbf{\Pi}$ is a block diagonal matrix defined by
\begin{eqnarray}
\mathbf{\Pi}  & \triangleq & \left[ \begin{array}{cccc}
\gamma_1 \mathbf{A}_1  &                        &           &    \\
                       & \gamma_2 \mathbf{A}_2  &           &  \\
                       &                        & \ddots    &  \\
                       &                        &           & \gamma_g \mathbf{A}_g \end{array} \right].  \label{equ:A_structure}
\end{eqnarray}
Besides, each row of the noise matrix $\mathbf{V}$ is assumed to have the distribution $p(\mathbf{V}_{i\cdot}; \lambda,\mathbf{B}) = \mathcal{N}(\mathbf{0},\lambda \mathbf{B})$, where $\lambda$ is an unknown scalar. Assuming the rows are mutually independent, we have
\begin{eqnarray}
p(\mathrm{vec}(\mathbf{V}^T); \lambda,\mathbf{B}) =  \mathcal{N}(\mathbf{0},\lambda \mathbf{I} \otimes \mathbf{B} ). \label{equ:V_assumption}
\end{eqnarray}

\textbf{Remark 1}: Note that $\mathbf{X}$ and $\mathbf{V}$ share the common matrix $\mathbf{B}$ for modeling the correlation structure of each row. This is a traditional setting in Bayesian variable selection models \cite{Brown1998multivariate}, which facilitates the use of conjugate priors for multivariate linear regression. Besides, since in our applications the sensor noise $\mathbf{V}$ can be ignored, the covariance model of $\mathbf{V}$ is not important. It only facilitates the development of our algorithm.

\textbf{Remark 2}: The proposed STSBL model is an extension of the model used by BSBL \cite{Zhang2012TSP}. Setting $L=1$, the STSBL model reduces to the latter. In other words, the STSBL model can be viewed as a set of multiple BSBL models with their solution vectors \emph{mutually correlated}. In Section \ref{subsec:interchannel} we will see the necessity of modeling the mutual correlation.

\textbf{Remark 3}: The proposed STSBL model is also closely related to the T-MSBL model \cite{Zhang2011IEEE} \footnote{Due to the difference in problem formulation, the temporal correlation studied in \cite{Zhang2011IEEE} is the inter-channel correlation in this work.}. When $d_i = 1 (\forall i)$, STSBL reduces to the latter. Note that T-MSBL only exploits  correlation among entries of the same row in $\mathbf{X}$, while STSBL also exploits  correlation among entries of the same column in $\mathbf{X}$. In the context of data compression, T-MSBL only exploits the inter-channel correlation, while STSBL exploits both the inter-channel correlation and the temporal correlation within each channel signal.

The relationships revealed  in Remark 2 and Remark 3 inspire us to derive an efficient algorithm, as shown below.


\section{The Spatiotemporal SBL Algorithm}
\label{sec:algorithm}

Due to the coupling between $\mathbf{A}_i(\forall i)$ and $\mathbf{B}$, directly estimating parameters in the model (\ref{equ:STmodel_original}) can result in an algorithm with heavy computational load. However, the observations in Remark 2 and Remark 3 imply that we can use $\mathbf{B}$ as a spatially whitening matrix, transforming the original model (\ref{equ:STmodel_original}) to a spatially whitened  model, and use $\mathbf{A}_i(\forall i)$ to transform the original model to a temporally whitened model \footnote{In fact, the block partition is still present. But for convenience, we call the equivalent model a ``temporally whitened'' model.}. Thus, we propose an \emph{alternating-learning approach}, where the parameters $\{\gamma_i,\mathbf{A}_i\}_{i=1}^g$ and $\lambda$ are estimated from the spatially whitened model, and the parameter $\mathbf{B}$ is estimated from the temporally whitened model. The resulting algorithm alternates the estimation between the two models until convergence. The alternating-learning approach largely simplifies the algorithm development.

\subsection{Learning in the Spatially Whitened Model}
\label{subsec:temporallywhitenedmodel}

To facilitate algorithm development, we assume $\mathbf{B}$ is known. Letting $\widetilde{\mathbf{Y}} \triangleq \mathbf{Y} \mathbf{B}^{-\frac{1}{2}}$, $\widetilde{\mathbf{X}} \triangleq \mathbf{X} \mathbf{B}^{-\frac{1}{2}}$, and $\widetilde{\mathbf{V}} \triangleq \mathbf{V} \mathbf{B}^{-\frac{1}{2}}$, the original STSBL model (\ref{equ:STmodel_original}) becomes
\begin{eqnarray}
\widetilde{\mathbf{Y}}= \mathbf{\Phi} \widetilde{\mathbf{X}} + \widetilde{\mathbf{V}},
\label{equ:model_temporallywhitened}
\end{eqnarray}
where the columns of $\widetilde{\mathbf{X}}$ are independent, and so does $\widetilde{\mathbf{V}}$. Thus, the original STSBL model is now spatially whitened, and the algorithm development becomes easier.

First, we have priors for $p(\widetilde{\mathbf{X}}; \mathbf{\Pi})$ and $p(\widetilde{\mathbf{V}}; \lambda)$ as follows
\begin{eqnarray}
p(\widetilde{\mathbf{X}};\mathbf{\Pi}) &=& \prod_{i=1}^L p(\widetilde{\mathbf{X}}_{\cdot i}) \sim \prod_i \mathcal{N}(\mathbf{0},\mathbf{\Pi})   \\
p(\widetilde{\mathbf{V}};\lambda) &=& \prod_{i=1}^L p(\widetilde{\mathbf{V}}_{\cdot i}) \sim \prod_i \mathcal{N}(\mathbf{0},\lambda \mathbf{I})
\end{eqnarray}
Then we have the likelihood:
\begin{eqnarray}
p(\widetilde{\mathbf{Y}}_{\cdot i} | \widetilde{\mathbf{X}}_{\cdot i}; \lambda) &=& \mathcal{N}(\mathbf{\Phi} \widetilde{\mathbf{X}}_{\cdot i}, \lambda \mathbf{I})  \quad \forall i
\end{eqnarray}
Thus, we obtain the posterior:
\begin{eqnarray}
p(\widetilde{\mathbf{X}}_{\cdot i} |\widetilde{\mathbf{Y}}_{\cdot i}; \lambda, \mathbf{\Pi}) &=& \mathcal{N}(\boldsymbol{\mu}_{\cdot i}, \mathbf{\Sigma}) \quad \forall i
\end{eqnarray}
with the mean $\boldsymbol{\mu}_{\cdot i}$ and the covariance matrix $\mathbf{\Sigma}$ given by
\begin{eqnarray}
\boldsymbol{\mu}_{\cdot i} &=& \mathbf{\Pi}\mathbf{\Phi}^T(\lambda \mathbf{I} + \mathbf{\Phi} \mathbf{\Pi} \mathbf{\Phi}^T)^{-1} \widetilde{\mathbf{Y}}_{\cdot i}  \quad \forall i \label{equ:temporalwhitenedmodel_mu}\\
\mathbf{\Sigma} &=& (\mathbf{\Pi}^{-1} + \frac{1}{\lambda} \mathbf{\Phi}^T \mathbf{\Phi})^{-1} \label{equ:temporalwhitenedmodel_Sigma} \\
&=& \mathbf{\Pi} - \mathbf{\Pi} \mathbf{\Phi}^T (\lambda \mathbf{I} + \mathbf{\Phi} \mathbf{\Pi} \mathbf{\Phi}^T)^{-1} \mathbf{\Phi} \mathbf{\Pi}  \label{equ:temporalwhitenedmodel_Sigma2}
\end{eqnarray}
Once the parameters $\mathbf{\Pi}$ and $\lambda$ are estimated, the maximum-a-posteriori (MAP) estimate of $\widetilde{\mathbf{X}}$ is directly given by the mean of the posterior, i.e.,
\begin{eqnarray}
\widetilde{\mathbf{X}} \leftarrow \mathbf{\Pi}\mathbf{\Phi}^T(\lambda \mathbf{I} + \mathbf{\Phi} \mathbf{\Pi} \mathbf{\Phi}^T)^{-1} \widetilde{\mathbf{Y}}, \label{equ:temporalwhitenedmodel_X}
\end{eqnarray}
and the solution matrix $\mathbf{X}$ in the original STSBL model (\ref{equ:STmodel_original}) can be obtained:
\begin{eqnarray}
\mathbf{X} \leftarrow \widetilde{\mathbf{X}} \mathbf{B}^{\frac{1}{2}}.
\end{eqnarray}

Thus, estimating the parameters $\mathbf{\Pi}$ and $\lambda$ is crucial to the algorithm. There are many optimization methods which can be used to estimate these parameters, such as bound-optimization methods \cite{Zhang2012TSP}, fast marginal likelihood maximization \cite{Tipping2003}, and variational methods \cite{Shutin2011fast}. In this work we use the expectation maximization (EM) method to estimate them, since we find the resulting algorithm can provide better recovery performance in our application.

Using the EM method, $\widetilde{\mathbf{X}}$ is treated as a hidden variable. The Q-function for estimating $\{\gamma_i\}_i$ and $\{\mathbf{A}_i\}_i$ is given by
\begin{eqnarray}
\mathcal{Q}(\mathbf{\Pi}) &\triangleq&  E_{\widetilde{\mathbf{X}}|\widetilde{\mathbf{Y}};\Theta^{(\mathrm{old})}} \big[\log p(\widetilde{\mathbf{X}};\{\gamma_i\}_i,\{\mathbf{A}_i\}_i) \big] \nonumber \\
&=& -\frac{L}{2} \log|\mathbf{\Pi}|  - \frac{1}{2} \sum_{i=1}^L E_{\widetilde{\mathbf{X}}|\widetilde{\mathbf{Y}};\Theta^{(\mathrm{old})}} \big[\widetilde{\mathbf{X}}_{\cdot i}^T \mathbf{\Pi}^{-1} \widetilde{\mathbf{X}}_{\cdot i}\big] \nonumber \\
&=& -\frac{L}{2} \sum_{i=1}^g \log |\gamma_i \mathbf{A}_i| \nonumber\\
&& -\frac{1}{2} \sum_{l=1}^L \mathrm{Tr}\Big[ \mathbf{\Pi}^{-1} \big( \mathbf{\Sigma} + \boldsymbol{\mu}_{\cdot l} \boldsymbol{\mu}_{\cdot l}^T  \big)\Big] \nonumber \\
&=& -\frac{L}{2} \sum_{i=1}^g d_i \log \gamma_i  - \frac{L}{2} \sum_{i=1}^g \log|\mathbf{A}_i| \nonumber \\
&& - \frac{1}{2} \sum_{l=1}^L \sum_{j=1}^g \frac{1}{\gamma_j} \mathrm{Tr}\Big[ \mathbf{A}_j^{-1} \big( \mathbf{\Sigma}_{[j]}  + \boldsymbol{\mu}_{[j] l} \boldsymbol{\mu}_{[j] l}^T  \big)\Big], \label{equ:Q4A}
\end{eqnarray}
where $\Theta^{(\mathrm{old})}$ denotes all the parameters estimated in the previous iteration, $\mathbf{\Sigma}_{[j]}$ denotes the $j$-th diagonal block in $\mathbf{\Sigma}$ which corresponds to the $j$-th block in $\widetilde{\mathbf{X}}$, and $\boldsymbol{\mu}_{[j] l}$ denotes the $j$-th block in the $l$-th column of $\boldsymbol{\mu}$.

Setting to zero the derivative of (\ref{equ:Q4A}) w.r.t. $\gamma_i$, we obtain the updating rule for $\gamma_i$:
\begin{eqnarray}
\gamma_i \leftarrow \frac{1}{Ld_i} \sum_{l=1}^L \mathrm{Tr}\Big[ \mathbf{A}_i^{-1} \big( \mathbf{\Sigma}_{[i]} + \boldsymbol{\mu}_{[i] l} \boldsymbol{\mu}_{[i] l}^T  \big)\Big]. \label{equ:STSBL_gamma}
\end{eqnarray}

Setting to zero the derivative of (\ref{equ:Q4A}) w.r.t. $\mathbf{A}_i$, we obtain the updating rule for $\mathbf{A}_i$:
\begin{eqnarray}
\mathbf{A}_i \leftarrow \frac{1}{L} \sum_{l=1}^L \frac{\mathbf{\Sigma}_{[i]} + \boldsymbol{\mu}_{[i] l} \boldsymbol{\mu}_{[i] l}^T}{\gamma_i}. \label{equ:STSBL_A}
\end{eqnarray}
The estimate will be further regularized as shown later.

To estimate $\lambda$, the Q-function is given by
\begin{eqnarray}
\mathcal{Q}(\lambda) &=&  E_{\widetilde{\mathbf{X}}|\widetilde{\mathbf{Y}};\Theta^{(\mathrm{old})}} \big[\log p(\widetilde{\mathbf{Y}}|\widetilde{\mathbf{X}};\lambda) \big] \nonumber \\
&\propto& -\frac{NL}{2} \log \lambda \nonumber \\
&& - \frac{1}{2\lambda} E_{\widetilde{\mathbf{X}}|\widetilde{\mathbf{Y}};\Theta^{(\mathrm{old})}} \Big[ \sum_{l=1}^L \| \widetilde{\mathbf{Y}}_{\cdot l} - \mathbf{\Phi} \widetilde{\mathbf{X}}_{\cdot l}  \|_2^2 \Big] \nonumber \\
&=& -\frac{NL}{2} \log \lambda - \frac{1}{2\lambda}\sum_{l=1}^L \Big[ \|\widetilde{\mathbf{Y}}_{\cdot l} - \mathbf{\Phi} \boldsymbol{\mu}_{\cdot l} \|_2^2   \nonumber \\
&& + E_{\widetilde{\mathbf{X}}|\widetilde{\mathbf{Y}};\Theta^{(\mathrm{old})}} \big[ \| \mathbf{\Phi} (\widetilde{\mathbf{X}}_{\cdot l} -  \boldsymbol{\mu}_{\cdot l} ) \|_2^2 \big] \Big] \nonumber \\
&=& -\frac{NL}{2} \log \lambda - \frac{1}{2\lambda} \|\widetilde{\mathbf{Y}} - \mathbf{\Phi} \boldsymbol{\mu} \|_\mathcal{F}^2 \nonumber \\
&& - \frac{L}{2\lambda} \mathrm{Tr}\big( \mathbf{\Sigma} \mathbf{\Phi}^T \mathbf{\Phi} \big).
\end{eqnarray}
Setting its derivative to zero, we have
\begin{eqnarray}
\lambda \leftarrow \frac{1}{NL} \|\widetilde{\mathbf{Y}} - \mathbf{\Phi} \boldsymbol{\mu} \|_\mathcal{F}^2 + \frac{1}{N}\mathrm{Tr}\big( \mathbf{\Sigma} \mathbf{\Phi}^T \mathbf{\Phi} \big). \label{equ:STSBL_lambda}
\end{eqnarray}
Similar to the approach adopted in \cite{Zhang2012TSP}, at low signal-to-noise (SNR) situations the above updating rule is modified to
\begin{eqnarray}
\lambda \leftarrow \frac{1}{NL} \|\widetilde{\mathbf{Y}} - \mathbf{\Phi} \boldsymbol{\mu} \|_\mathcal{F}^2 + \frac{1}{N} \sum_{i=1}^g \mathrm{Tr}\big( \mathbf{\Sigma}_{[i]} \mathbf{\Phi}_{\cdot [i]}^T \mathbf{\Phi}_{\cdot [i]} \big), \label{equ:STSBL_lambda_lowSNR}
\end{eqnarray}
where $\mathbf{\Phi}_{\cdot [i]}$ denotes the consecutive columns in $\mathbf{\Phi}$ which correspond to the $i$-th block in $\widetilde{\mathbf{X}}$. In noiseless situations one can simply set $\lambda=10^{-10}$ or other small values, instead of performing the updating rule (\ref{equ:STSBL_lambda}).

In the above development we have assumed that $\mathbf{B}$ is given. This parameter can be estimated in a temporally whitened model discussed below.

\subsection{Learning in the Temporally Whitened Model}

To estimate the matrix $\mathbf{B}$, we consider the following equivalent form of the original model (\ref{equ:STmodel_original}):
\begin{eqnarray}
\mathbf{Y} = \overline{\mathbf{\Phi}} \cdot \overline{\mathbf{X}} + \mathbf{V} \label{equ:model_spatiallywhitened}
\end{eqnarray}
where $\overline{\mathbf{\Phi}} \triangleq \mathbf{\Phi} \mathbf{A}^{\frac{1}{2}}$, $\overline{\mathbf{X}} \triangleq \mathbf{A}^{-\frac{1}{2}}  \mathbf{X}$, and $\mathbf{A}$ is defined as $\mathbf{A} \triangleq \mathrm{diag}\{\mathbf{A}_1,\cdots,\mathbf{A}_g\}$. Note that in this model, $\overline{\mathbf{X}}$ maintains the same block partition as $\mathbf{X}$, but its every block has no temporal correlation due to the temporally whitening effect from $\mathbf{A}_i^{-\frac{1}{2}}(\forall i)$. Thus, estimating $\mathbf{B}$ in this model becomes easier.

Following the approach used to derive the T-MSBL algorithm \cite{Zhang2011IEEE} and assuming $\mathbf{X}$, $\{\gamma_i\}_i$ and $\{\mathbf{A}_i\}_i$ have been obtained from the spatially whitened model (\ref{equ:model_temporallywhitened}), we have the following updating rule for the matrix $\mathbf{B}$:
\begin{eqnarray}
\check{\mathbf{B}} & \leftarrow & \sum_{i=1}^{g}  \gamma_i^{-1} \overline{\mathbf{X}}_{[i] \cdot}^T    \overline{\mathbf{X}}_{[i] \cdot} + \lambda^{-1} (\mathbf{Y}-\overline{\mathbf{\Phi}} \overline{\mathbf{X}} )^T(\mathbf{Y}-\overline{\mathbf{\Phi}} \overline{\mathbf{X}}) \nonumber \\
&=&  \sum_{i=1}^{g}  \frac{  \mathbf{X}_{[i] \cdot}^T \mathbf{A}_i^{-1}   \mathbf{X}_{[i] \cdot} }{\gamma_i} + \frac{   (\mathbf{Y}-\mathbf{\Phi} \mathbf{X} )^T(\mathbf{Y}-\mathbf{\Phi} \mathbf{X}) }{\lambda} \label{equ:STSBL_B}\\
\mathbf{B} & \leftarrow & \check{\mathbf{B}}/\|\check{\mathbf{B}}\|_\mathcal{F} \label{equ:STSBL_B_normalize}
\end{eqnarray}
where $\overline{\mathbf{X}}_{[i] \cdot} \in \mathbb{R}^{d_i \times L}$ is the $i$-th block in $\overline{\mathbf{X}}$, and  $\overline{\mathbf{X}}_{[i] \cdot} \triangleq \mathbf{A}_i^{-\frac{1}{2}}  \mathbf{X}_{[i] \cdot}$.
The second item in (\ref{equ:STSBL_B}) is noise-related. When the noise is very small, or does not exist (i.e., $\lambda \rightarrow 0$), it is
suggested to remove the second item for robustness.

\subsection{Regularization}
\label{subsec:regularization}

In the proposed spatiotemporal model the number of unknown parameters is much larger than the number of available data. Thus regularization to the estimated $\mathbf{B}$ and $\{\mathbf{A}_i\}_i$ is very important. Suitable regularization helps to overcome learning difficulties resulting from the very limited data.

As in \cite{Zhang2011IEEE}, we can regularize the $\check{\mathbf{B}}$ in (\ref{equ:STSBL_B}) by
\begin{eqnarray}
\check{\mathbf{B}} \leftarrow   \sum_{i=1}^{g}  \gamma_i^{-1} \mathbf{X}_{[i] \cdot}^T \mathbf{A}_i^{-1}   \mathbf{X}_{[i] \cdot} + \eta \mathbf{I} \label{equ:STSBL_B_reg}
\end{eqnarray}
where $\eta$ is a positive scalar. This regularization is shown empirically to increase robustness in noisy environments. In noiseless environments, this regularization is not needed.

To regularize the estimates of $\{\mathbf{A}_i\}_i$, we use the strategy in \cite{Zhang2012TSP}, i.e., modeling the correlation matrix of each column in $\mathbf{X}_{[i]\cdot}$ as the correlation matrix of an AR(1) process with the common AR coefficient $r$ for all $i$. The strategy can be summarized as follows.
\begin{itemize}
  \item Step 1: From each $\mathbf{A}_i$, calculate the quantity $r_i$ by $r_i  \leftarrow    \frac{m_1^i}{m_0^i}\, (\forall i)$, where $m_0^i$ is the average of entries in the main diagonal of $\mathbf{A}_i$ and $m_1^i$ is the average of entries in the main sub-diagonal of $\mathbf{A}_i$. Note that due to some numerical problems, $\frac{m_1^i}{m_0^i}$ may be out of the feasible range $(-1,1)$, and thus further constraints may be imposed; for example, $r_i \leftarrow \mathrm{sign}(\frac{m_1^i}{m_0^i} ) \min\{|\frac{m_1^i}{m_0^i} |,0.99\}$;

  \item Step 2: Average: $r  \leftarrow   \frac{1}{g}\sum_{i=1}^g r_i$

  \item Step 3: Reconstruct the regularized $\mathbf{A}_i (\forall i)$:
          \begin{eqnarray}
            \check{\mathbf{A}}_i  &\leftarrow & \left[
                \begin {array}{cccc}
                1             & r                     & \cdots   & r^{d_i-1} \\
                \vdots        &  \vdots       &                 & \vdots     \\
                r^{d_i-1}       & r^{d_i-2}       &  \cdots               & 1
            \end {array} \right] \nonumber \\
            \mathbf{A}_i  &\leftarrow & \check{\mathbf{A}}_i/\|\check{\mathbf{A}}_i \|_\mathcal{F}  \nonumber
          \end{eqnarray}

\end{itemize}

The parameter-averaging strategy has been widely used in artificial neural networks and the machine learning communities to overcome overfitting.

Experiments showed these regularization strategies helped further improve the algorithm's performance. In fact, using the Theorem 1 in \cite{Zhang2011IEEE} it can be proved that in noiseless situations the regularization strategies to $\mathbf{A}_i$ and $\mathbf{B}$ do not affect the global minimum of the cost function of our algorithm, in the sense that the global minimum corresponds to the true sparse solution. This implies that a good regularization strategy can significantly enhance global convergence of our algorithm.

Up to now we have derived the updating rules for $\mathbf{X}$, $\{\mathbf{A}_i\}_i$, $\{\gamma_i\}_i$ and $\lambda$ in the spatially whitened model and the updating rules for $\mathbf{B}$ in the temporally whitened model. Combining these updating rules we obtain the EM-based spatiotemporal sparse Bayesian learning algorithm, denoted by \textbf{STSBL-EM}.

\section{Practical Considerations When Applying STSBL-EM}
\label{sec:apply2CS}

The proposed STSBL-EM algorithm has wide applications. This section discusses some practical considerations when applying it in practice.

In CS of multichannel physiological signals, if the channel signal $\mathbf{X}_{\cdot l}$ has strong temporal correlation \footnote{Here `strong temporal correlation' means that if modeling the signal by an AR(1) process, the absolute value of the AR coefficient is very large.} in the time domain, using the original spatiotemporal model (\ref{equ:STmodel_original}) can achieve good recovery performance. A typical signal is ECG signals \cite{Zhang2012TBME}.

When each channel signal $\mathbf{X}_{\cdot l}$ does not have strong temporal correlation, exploiting the temporal correlation may not be very beneficial. Thus one can alternatively exploit the sparsity of each channel signal in some transformed domain by using a dictionary matrix in STSBL-EM, as stated in Section \ref{sec:intro}. In particular, one can first apply the algorithm  to the following model
\begin{eqnarray}
\mathbf{Y} = \mathbf{\Omega} \mathbf{Z} + \mathbf{V}
\label{equ:model_spatiotemporal_withD}
\end{eqnarray}
to find the solution $\mathbf{Z}$, where $\mathbf{\Omega} \triangleq \mathbf{\Phi} \mathbf{D}$, and $\mathbf{D}$ is a dictionary matrix under which $\mathbf{X}_{\cdot l} (\forall l)$ has sparse representation $\mathbf{Z}_{\cdot l}$. Then one can obtain the original solution by computing $\mathbf{X} = \mathbf{DZ}$. Note that in this method $\mathbf{Z}_{\cdot l}$ is sparser than $\mathbf{X}_{\cdot l}$, but generally has less correlation than the latter, or the correlation structure in $\mathbf{Z}_{\cdot l}$ is not well captured by STSBL-EM. Hence, this method mainly exploits each channel signal's sparsity in a transformed domain instead of exploiting the channel signal's temporal correlation \footnote{Note that when using some dictionary matrices such as wavelet dictionaries, one may exploit both  sparsity and wavelet tree structures in $\mathbf{Z}_{\cdot l}$, which is more beneficial than merely exploiting the sparsity \cite{he2009exploiting,chen2012compressive}.}. This method can yield better results than using the original model (\ref{equ:STmodel_original}), if each channel signal has no strong temporal correlation. A typical signal is EEG signals \cite{Zhang2012Letter}.

In the following experiments on EEG signals we will adopt the model (\ref{equ:model_spatiotemporal_withD}) with the dictionary matrix $\mathbf{D}$ formed by the orthogonal DCT bases \footnote{One may find other dictionary matrices which can yield better results than the DCT dictionary matrix on EEG signals \cite{Abdulghani2012EEG}. But seeking the optimal dictionary matrix is not the focus of this work.}. Due to the ``energy compaction" property of DCT, for the $l$-th channel signal $\mathbf{X}_{\cdot l}$, the DCT coefficients with significantly nonzero values are concentrated in the first $K$ entries in $\mathbf{Z}_{\cdot l}$. Note that the first $K$ nonzero entries (with other  coefficients with insignificantly nonzero values locating at the $(K+1)$-th entry, the $(K+2)$-th entry, etc.) can be viewed as  concatenation of a number of nonzero blocks. In this sense, the value of $K$ does not need to be known \emph{a priori}, and the block partition in STSBL-EM can be set rather arbitrarily.
In our experiments we found STSBL-EM showed stable performance when the block partition $d_i (\forall i)$ chose values from a wide range (15 to 60). (Similar robustness was also observed on BSBL \cite{Zhang2012TBME}.) Thus we simply set $d_i = 16 (\forall i)$.

In practice most SBL algorithms implicitly adopt a $\gamma_i$-pruning mechanism \cite{Zhang2011IEEE,Zhang2012TSP,Tipping2001}. The mechanism sets a small $\gamma_i$ to zero if it is smaller than a threshold, thus speeding up convergence and encouraging solutions to be sparse in the level of entries \cite{Tipping2001}, blocks \cite{Zhang2012TSP}, or rows \cite{Zhang2011IEEE}. However, for raw EEG signals (especially those recorded during ambulatory monitoring) the value of $K/M$ could be very large \cite{Zhang2013Asilomar}. Thus the DCT coefficient vectors are not sparse. In this case, better recovery performance can be achieved by setting the $\gamma_i$-pruning threshold to a very small value or even zero and allowing algorithms to iterate only a few times \cite{Zhang2012TBME,Zhang2012Letter}. In our experiments we set this threshold to zero, and terminated the algorithm when the iteration number reached 40 or the maximum change in any entry of the estimated $\mathbf{X}$ in two successive iterations was smaller than $10^{-6}$. But when used in other applications such as source localization, it may need hundreds of iterations to converge.

In our work the problem of data compression is modeled as a noiseless CS problem (i.e., the sensor noise $\mathbf{V}$ is ignored). Therefore, in our experiments STSBL-EM was performed in the noiseless situation with the parameter $\lambda$ set to $\lambda=10^{-10}$. But this does not mean that  artifacts and noise in raw physiological signals are ignored. In fact, in our model $\mathbf{X}_{\cdot l}$ is a raw physiological signal contaminated by noise and artifacts.

\section{Application}
\label{sec:experiments}

The proposed STSBL-EM was used for CS of multichannel EEG signals  in SSVEP-based BCI and EEG-based driver's drowsiness estimation.

To show the superior performance of STSBL-EM, we chose the BSBL-BO algorithm, an MMV-model-based CS algorithm, and an SMV-model-based CS algorithm  for comparison. We did not choose many algorithms for comparison, since in \cite{Zhang2012TBME} it has been shown that ten state-of-the-art CS algorithms were inferior to BSBL-BO. Thus, our focus was the comparison between STSBL-EM and BSBL-BO. The three algorithms are briefly described as follows.
\begin{itemize}
  \item BSBL-BO \cite{Zhang2012TSP} \footnote{The Matlab code was downloaded at \url{https://sites.google.com/site/researchbyzhang/bsbl}.}. To the best of our knowledge, it may be the only algorithm that has the ability to recover both non-sparse physiological signals \cite{Zhang2012TBME} and the physiological signals with non-sparse representation coefficients \cite{Zhang2012Letter}. Its block partition was set to ${d_1=d_2=\cdots = 16}$.

  \item ISL0 \cite{Hyder2010b} \footnote{The Matlab code was provided by the first author of \cite{Hyder2010b} via private communication.}. It is based on the MMV model. When $\mathbf{Z}$ is less row-sparse, it has robust performance than many MMV-model-based algorithms.

  \item Basis Pursuit (BP) \cite{BP} \footnote{The Matlab code was downloaded at \url{http://www.cs.ubc.ca/~mpf/spgl1/}.}. It is a classic CS algorithm based on the SMV model. Some work \cite{Abdulghani2012EEG} claimed that it was more suitable for CS of EEG than other classic CS algorithms. We used the SPGL1 software \cite{van2008probing} to implement this algorithm.
\end{itemize}

All the algorithms recovered signals in the transformed domain. The dictionary matrix $\mathbf{D}$ was the DCT dictionary matrix. For all algorithms, the measurement matrix $\mathbf{\Phi}$ was an $N \times M$ sparse binary matrix of full row-rank, where $M$ was fixed to 256 and $N$ was varied to meet a desired compression ratio (CR). The CR was defined as
\begin{eqnarray}
CR = \frac{M-N}{M} \times 100.
\end{eqnarray}
Irrespective of  CR values, each column of the measurement matrix $\mathbf{\Phi}$ contained only \emph{two} entries of 1's with random locations, while other entries were zeros.

Mean square error (MSE) is often used for measuring recovery quality. However, it is shown \cite{Wang2009MSE} that MSE is not a reasonable measure for natural signals. Thus it is not suitable for EEG, especially raw EEG signals contaminated by strong noise and artifacts. A smaller MSE does not necessarily mean that a desired task (e.g. EEG classification) on the recovered EEG signals can be better accomplished. Therefore, we used a task-oriented performance evaluation method, which was initially suggested in \cite{Zhang2012TBME,Zhang2012Letter}.

The main idea of this evaluation method is that a practical task is first performed on an original dataset, and then the same task (using the same algorithm with the same initialization) is performed on the recovered dataset, and finally the results of the two tasks are compared. If the results are the same,  this means that the recovered dataset has high fidelity and does not affect the practical task. If the results are far from each other, this means that the recovered dataset is seriously distorted.

Using this idea, in our BCI experiment we compared the classification rate on original EEG signals to the classification rate on  recovered signals. In the experiment on drowsiness estimation, we compared the estimation result using original signals to the estimation result using recovered signals. All the comparisons were repeated with different CR values.

Experiments were carried out on a computer with dual-core 2.8 GHz CPU and 6.0 GiB RAM.

\subsection{SSVEP-Based BCI}

In neurology, SSVEP is a response to a visual stimulus modulated at a specific frequency. The response has a fundamental frequency  and several harmonics. The fundamental frequency is the same as that of the visual stimulus. This characteristic has been widely used in BCI \cite{Wang2008SSVEP} to classify stimuli with different frequencies, thereby finishing some control tasks. A trend in BCI is to develop wearable wireless systems \cite{Chi2012dry,liao2012gaming}. In such systems developing energy-efficient data acquisition modules is highly desired.

In this experiment the dataset analyzed in \cite{Chi2012dry}  was used \footnote{The dataset was downloaded at \url{ftp://sccn.ucsd.edu/pub/SSVEP}.}. The dataset was recorded from twelve subjects. We chose the recordings of `Subject 1' for illustration, which corresponded to visual stimuli of 9Hz, 10Hz, 11Hz, 12Hz, and 13Hz. Each stimulus flashed for 4 seconds. The data sampling rate was 256 Hz. The monitor refresh rate was 75Hz. As in \cite{Chi2012dry}, canonical correlation analysis (CCA) was used as the classifier. The selected channel indexes were 129, 133, 193, 196, 199, 200, 203, and 210 (all in the occipital area). Detailed descriptions on the dataset, the experiment equipment, and the recording procedure can be found in \cite{Chi2012dry}.

The signals were compressed and then recovered by STSBL-EM, BSBL-BO, ISL0, and BP, respectively. CR ranged from 50 to 90. The recovered signals were  bandpass-filtered between 8-35 Hz. Each 8-channel epoch which corresponded to a visual stimulus was classified by CCA. The classification rate was calculated by averaging over all classification results on the whole recovered signals. The same bandpass filtering and classification were performed on the original signals.

The classification rates of all algorithms are given in Table \ref{table:app6_SSVEP4}. Note that the classification rate on the original signals was 1.00. We can see that when $\mathrm{CR} \leq 60$, the classification rate on the recovered signals by STSBL-EM was also 1.00. Even if $\mathrm{CR} = 80$, the classification rate was very close to 1.00. These results imply that when the signals were compressed by 80\%, the recovered signals by our algorithm were still of good quality. In contrast, all the compared algorithms did not recover the signals with satisfactory quality even with $\mathrm{CR}=60$.

\begin{table}[tp]
\renewcommand{\arraystretch}{1.2}
\caption{Classification rates of all algorithms when CR varied from 50 to 90. The classification rate on the original signals was \textbf{1.00}.}
\label{table:app6_SSVEP4}
\centering
\begin{tabular}{ccccccc}
\toprule
CR            &  50           &  60   & 70   & 80   & 85  & 90     \\
\midrule
STSBL-EM     &   \textbf{1.00}  &   \textbf{1.00} & \textbf{0.984}  & \textbf{0.984}   & \textbf{ 0.976}  & \textbf{0.672}  \\
\hline
BSBL-BO      &  0.992&  0.976   & 0.952  & 0.864   & 0.824   & 0.576  \\
\hline
ISL0         &  0.888&  0.840            & 0.800   & 0.704  & 0.536  & 0.488 \\
\hline
Basis Pursuit & 0.944& 0.920     & 0.856 & 0.728 & 0.600 & 0.528 \\
\bottomrule
\end{tabular}
\end{table}

\begin{figure}[tp]
\centering
\includegraphics[width=8cm,height=7cm]{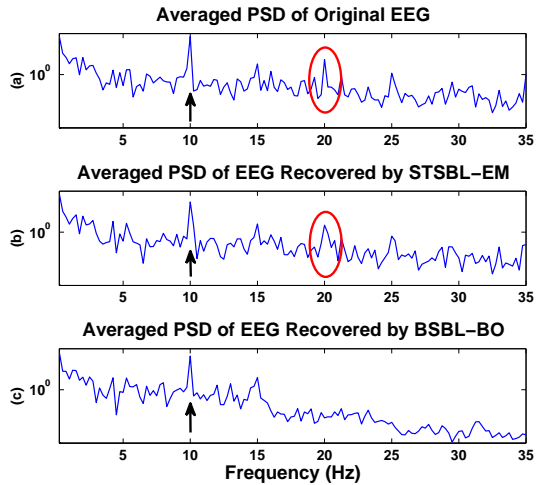}
\caption{(a) Averaged PSD of signals from the original signals, which corresponded to a visual stimulus of 10 Hz. (b) Averaged PSD of signals from the signals recovered by STSBL-EM when $\mathrm{CR}=80$. (c) Averaged PSD of signals from the signals recovered by BSBL-BO when $\mathrm{CR}=80$. Arrows indicate the fundamental frequency (10 Hz). Circles indicate the harmonic frequency (20 Hz).}
\label{fig:app6_10hz}
\end{figure}

To visually examine the data recovery quality, we randomly chose a time slot which corresponded to a visual stimulus of 10 Hz (duration was 4 seconds). Then we picked signals during this time slot in each channel from the original signals, and averaged their power spectrum densities (PSD's), shown in Fig.~\ref{fig:app6_10hz}(a). We can clearly see the fundamental frequency (10 Hz) and the harmonic frequency (20 Hz). Similarly, we calculated the averaged PSD from the recovered signals by STSBL-EM when $\mathrm{CR}=80$, shown in Fig.~\ref{fig:app6_10hz}(b), and the averaged PSD from the recovered signals by BSBL-BO when $\mathrm{CR}=80$, shown in Fig.~\ref{fig:app6_10hz}(c). We can  see both the fundamental frequency and the harmonic frequency in Fig.~\ref{fig:app6_10hz}(b). But we do not see the harmonic frequency in Fig.~\ref{fig:app6_10hz}(c). This explains why the classification rate on the signals recovered by BSBL-BO was lower than the classification rate on the signals recovered by STSBL-EM, since CCA exploited both the fundamental frequency and the harmonic frequency for classification.

Maintaining harmonic frequencies on recovered signals implies subtle waveforms in original signals are recovered. Therefore, the results shown in Fig. \ref{fig:app6_10hz} further confirms that STSBL-EM has better data recovery quality than BSBL-BO.

\begin{figure}[t]
\centering
\includegraphics[width=7cm,height=5.5cm]{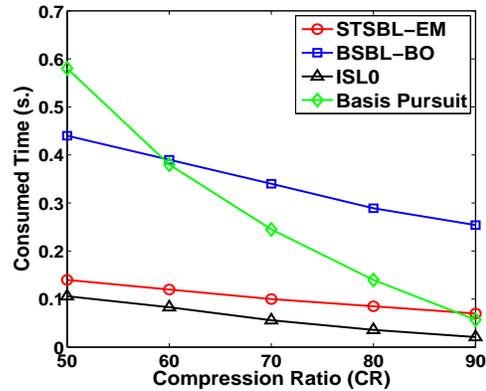}
\caption{Comparison of consumed time in recovery of 8-channel signals of 1 second duration at different CR. Only when the consumed time of an algorithm is far less than 1 second, can it be used for real-time (or near real-time) systems.}
\label{fig:app6_speed}
\end{figure}

Fig. \ref{fig:app6_speed} shows the averaged consumed time of each algorithm in recovering 8-channel signals of 1 second duration at different CR values. STSBL-EM was much faster than BSBL-BO. Their speed gap will be more significant in the next application, in which the number of EEG channels was 30.

\subsection{EEG-Based Driver's Drowsiness Estimation}

EEG-based driver's drowsiness estimation and prediction is an emerging technology for driving safety \cite{Jung1997estimating,lin2008noninvasive,Lin2005eeg} and an important application of EEG. Such systems are powered by batteries and are generally embedded in a wearable device such as an ordinary hat. Thus it is highly desired to develop wireless EEG systems with low energy consumption \cite{lin2008noninvasive}. In the following we will show that the proposed algorithm can be used in this application for energy efficient data transmission.

A set of EEG signals used in \cite{lin2008noninvasive} were used in this experiment. The data were recorded from a  subject using a 30-channel  EEG system, when the subject was driving with some degree of drowsiness in a realistic kinesthetic virtual-reality driving simulator. The sampling rate was 250 Hz. During the driving, the deviation between the center of the vehicle and the center of the cruising lane was recorded, which was viewed as a driving error. The driving error is known to be a good indicator to drowsiness level \cite{lin2008noninvasive,Lin2005eeg}. Details on the recording system, the recording procedure, and the virtual-reality driving simulator are given in \cite{lin2008noninvasive}.

Many methods were proposed to estimate the drowsiness level from recorded EEG signals. One method is given in \cite{Lin2005eeg,lin2008noninvasive}:
\begin{itemize}
  \item Use lowpass filter with a cut-off frequency of 50 Hz to remove power line noise and other high-frequency noise from raw EEG signals.
  \item Perform online independent component analysis (ICA) \cite{Lee1999} on the signals, and select an independent component (IC) for further analysis.
  \item Calculate log PSD of the selected IC at a frequency every 2 seconds. The time-varying subband log PSD is then used as the drowsiness estimate \footnote{For more robust estimation, one can seek an optimal mapping from the log PSD to the driving error using a training set. Since our goal in this experiment was to show the data recovery quality of the proposed algorithm, we just simply treated the time-varying log PSD as the drowsiness estimate.}.
\end{itemize}
To evaluate the quality of the drowsiness estimate,  the Pearson correlation between the driving error (an indicator to the drowsiness level) and the time-varying subband log PSD of the selected IC is often evaluated. High Pearson correlation indicates a good drowsiness estimate. Details of the method can be found in \cite{Lin2005eeg}.

Since our goal is to show that the proposed algorithm can be used in this application, we need to investigate whether the drowsiness estimation accuracy is degraded when using the recovered signals. Thus, we compared the drowsiness estimate from the recovered signals to the one from the original signals. Particularly, we adopted the following procedure:
\begin{enumerate}
  \item Repeat the above drowsiness estimation using the original signals by selecting an IC (denoted by $\mathrm{IC}_0$) and a frequency $f$. Evaluate the Pearson correlation between the driving error and the time-varying log PSD of $\mathrm{IC}_0$ at the frequency $f$. Denote the correlation by $r_0$.

  \item Perform the same ICA decomposition on the recovered signals, and choose the IC which has the highest Pearson correlation with $\mathrm{IC}_0$. Denote the IC by $\mathrm{IC}_{\mathrm{rec}}$.

  \item Calculate the time-varying log PSD of $\mathrm{IC}_{\mathrm{rec}}$ at the frequency $f$.

  \item Evaluate the Pearson correlation between the driving error and the time-varying log PSD calculated in the above step. Denote the Pearson correlation by $r_{\mathrm{rec}}$.

  \item Compare $r_{\mathrm{rec}}$ to $r_0$.
\end{enumerate}
In our experiment,  $\mathrm{IC}_0$ was the IC whose log PSD at $f=5$ (Hz) had the highest correlation with the driving error.

\begin{figure}[t]
\centering
\includegraphics[width=9cm,height=9cm]{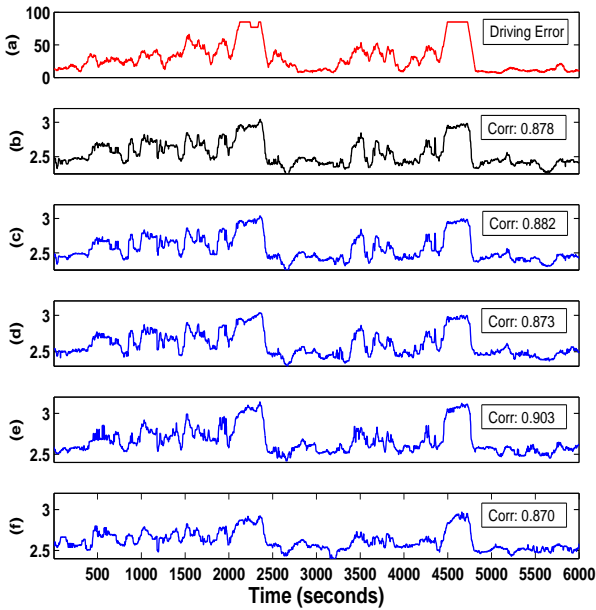}
\caption{Comparison of the driving error, the log PSD of $\mathrm{IC}_0$ at $f=5$ Hz, and the log PSD of $\mathrm{IC}_\mathrm{rec}$ at $f=5$ Hz and at different CR values. $\mathrm{IC}_\mathrm{rec}$ was obtained from recovered signals by STSBL-EM which had the highest correlation with $\mathrm{IC}_0$. (a) The driving error. (b) The log PSD of  $\mathrm{IC}_0$ at $f=5$ Hz. (c)-(f) are the log PSD of $\mathrm{IC}_\mathrm{rec}$ at $f=5$ Hz when CR = 50, 60, 70, and 80, respectively. Their Pearson correlations with the driving error are shown in each subplot.}
\label{fig:app2_PSD}
\end{figure}

Fig. \ref{fig:app2_PSD} shows the driving error signal, the time-varying log PSD of $\mathrm{IC}_0$ at $f=5$ Hz, and the time-varying log PSD of $\mathrm{IC}_{\mathrm{rec}}$ at $f=5$ Hz at different CR values. $\mathrm{IC}_\mathrm{rec}$ was obtained from recovered signals by STSBL-EM. The $r_0$ and the $r_{\mathrm{rec}}$ at different CR values are also given in corresponding subplots. Clearly, when CR was no more than 80, the drowsiness estimate from the recovered signals by STSBL-EM was almost the same as the one from the original signals.

Table \ref{table:app2_PC} further shows the $r_0$ and the $r_{\mathrm{rec}}$'s of all algorithms when $f=4,5,6,7$ (Hz) and CR varied from 50 to 80. We can see when CR was small (e.g. 50-60), all the algorithms recovered the signals well. Their drowsiness estimates were almost the same as the estimate from the original signals. However, when CR increased, only STSBL-EM ensured accurate drowsiness estimation; particularly, the drowsiness estimate was almost not affected even if the raw EEG signals were compressed by 80\%.

\begin{table*}[tp]
    \centering
    \renewcommand{\arraystretch}{1.2}
\caption{Comparison between $r_0$ calculated from the original signals and $r_{\mathrm{rec}}$  calculated from recovered signals by all algorithms at 4-7 Hz and different CR values. `--' means the ICA decomposition on the recovered signals by the corresponding algorithm did not yield the desired IC.}
\label{table:app2_PC}
    \begin{tabular}{@{}rrrrrcrrrr@{}}\toprule
        & \multicolumn{4}{c}{4Hz} & \phantom{x} & \multicolumn{4}{c}{5Hz} \\
        \cmidrule{2-5} \cmidrule{7-10}
        $\mathrm{CR}$ & $50$ & $60$ & $70$ & $80$ && $50$ & $60$ & $70$ & $80$ \\ \midrule
        Original Data & \multicolumn{4}{c}{$0.853$} && \multicolumn{4}{c}{$0.878$} \\
        STSBL-EM & $\mathbf{0.853}$ & $\mathbf{0.842}$ & $\mathbf{0.866}$ & $\mathbf{0.848}$ && $\mathbf{0.882}$ & $\mathbf{0.873}$ & $\mathbf{0.903}$ & $\mathbf{0.870}$ \\
        BSBL-BO & $0.853$ & $0.841$ & $0.841$ & $0.793$ && $0.880$ & $0.875$ & $0.874$ & $0.776$ \\
        ISL0 & $0.851$ & $0.735$ & - & - && $0.885$ & $0.776$ & - & - \\
        Basis Pursuit & $0.839$ & $0.842$ & $0.824$ & $0.780$ && $0.873$ & $0.854$ & $0.840$ & $0.795$ \\ \midrule
        & \multicolumn{4}{c}{6Hz} & \phantom{x} & \multicolumn{4}{c}{7Hz} \\
        \cmidrule{2-5} \cmidrule{7-10}
        $\mathrm{CR}$ & $50$ & $60$ & $70$ & $80$ && $50$ & $60$ & $70$ & $80$ \\ \midrule
        Original Data & \multicolumn{4}{c}{$0.881$} && \multicolumn{4}{c}{$0.807$} \\
        STSBL-EM & $\mathbf{0.879}$ & $\mathbf{0.870}$ & $\mathbf{0.896}$ & $\mathbf{0.867}$ && $\mathbf{0.809}$ & $\mathbf{0.771}$ & $\mathbf{0.849}$ & $\mathbf{0.808}$ \\
        BSBL-BO & $0.873$ & $0.871$ & $0.882$ & $0.733$ && $0.788$ & $0.801$ & $0.802$ & $0.526$ \\
        ISL0 & $0.874$ & $0.783$ & - & - && $0.806$ & $0.654$ & - & - \\
        Basis Pursuit & $0.867$ & $0.842$ & $0.808$ & $0.766$ && $0.779$ & $0.753$ & $0.728$ & $0.584$ \\
        \bottomrule
    \end{tabular}
\end{table*}

\begin{figure}[t]
\centering
\includegraphics[width=7cm,height=5.5cm]{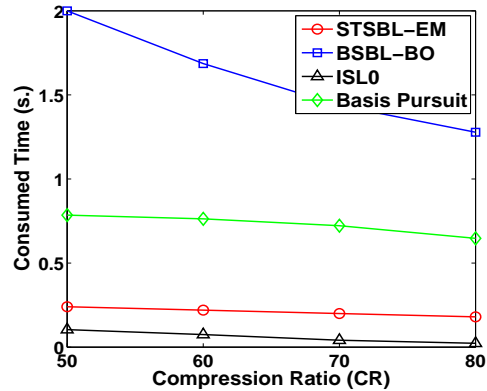}
\caption{Averaged consumed time of all algorithms in recovery of the 30-channel signals of 1.024 second duration at different CR values. BSBL-BO was slow, because it had to recover these signals channel by channel.}
\label{fig:app2_speed}
\end{figure}

Fig. \ref{fig:app2_speed} shows the averaged consumed time of all algorithms in recovery of the 30-channel signals of 1.024 second duration at different CR values. STSBL-EM was much faster than BSBL-BO, suggesting that it is more suitable for real-time applications especially when the channel number is very large.

It is worth pointing out that the raw EEG signals contained strong artifacts due to muscle movement. However, the proposed algorithm did not require any preprocessing before data compression.

\section{Discussions}
\label{sec:discussion}

\subsection{Energy Consumption}

We  have mentioned that the proposed algorithm compresses data with
ultra-low energy consumption.  This is due to the use of the simplest
measurement matrix and the algorithm's powerful data recovery ability.

The measurement matrix $\mathbf{\Phi}$ is a very simple sparse binary matrix.
Its each column contains only two entries of 1's, while other entries
are zeros. Using  this matrix has two major benefits,
\begin{itemize}
  \item Code execution in data compression is largely reduced. Consequently, the energy dissipated in code execution is very low.

  \item Using this measurement matrix largely
simplifies circuit design. Therefore the cost and the size of
chips can be reduced.
\end{itemize}

It is worth noting that such a measurement matrix is not suitable for
any CS algorithms. Some algorithms may have seriously degraded performance
when using the measurement matrix.

Besides, many CS algorithms require preprocessing on raw data before
compression, such as dynamic thresholding, filtering, and seeking
specific waveform features. These preprocessing consumes lots of energy
\footnote{It is highly doubted that if using such preprocessing, CS still has its energy-saving advantages over traditional data compression algorithms. }. In contrast, our proposed
algorithm does not require these preprocessing steps.

On the other hand, our algorithm's powerful recovery ability ensures high recovery performance
when the compression ratio is high (e.g. CR=80).  Thus, the energy
dissipated in wireless transmission can also be largely reduced.

In \cite{LiuZhang2013,fauvel2014energy} the compression procedure of BSBL-BO was analyzed. These works showed that BSBL-BO, compared to conventional data compression procedures, dissipated only about 10\% to 20\% energy, shortened compression time by more than 90\%, and largely saved other computational resources. Since the compression procedures of BSBL and STSBL-EM are the same, these analysis results are applicable to STSBL-EM.  But it is worth noting that STSBL-EM has more powerful recovery ability than BSBL-BO.

\subsection{Stable Speed Regardless of Channel Numbers}

Comparing Fig. \ref{fig:app2_speed} with Fig. \ref{fig:app6_speed} we find that the consumed time of STSBL-EM  was relatively stable, although the channel number in Fig. \ref{fig:app2_speed} was almost four times of the channel number in Fig. \ref{fig:app6_speed}. The reason is that to recover multichannel physiological signals, the algorithmic complexity of STSBL-EM mainly depends on the computation of (\ref{equ:temporalwhitenedmodel_Sigma2}) and (\ref{equ:temporalwhitenedmodel_X}), which is totally $\mathcal{O}(M^3 + 2 M^2 N + MN^2 + MNL + N^3)$. When $L$ is small compared to $M$ \footnote{In a typical scenario of telemonitoring of multichannel physiological signals, $L$ varies from two to dozens, while $M$ varies from 200 to 1000.}, the algorithmic complexity is approximately $\mathcal{O}(M^3 + 2 M^2 N)$, which does not depends on $L$. Thus the consumed time of STSBL-EM does not change significantly when the channel number dramatically changes. Note that when recovering a single-channel signal, the algorithmic complexity of BSBL-BO is also dominated by $\mathcal{O}(M^3 + 2 M^2 N)$. But when recovering $L$-channel signals, its computation load increases $L$-fold, since it has to recover the signals channel by channel. This explains why the consumed  time  by BSBL-BO in Fig. \ref{fig:app2_speed}  was roughly four times the consumed  time in Fig. \ref{fig:app6_speed}.

\subsection{Exploitation of Inter-Channel Correlation}
\label{subsec:interchannel}

Jointly recovering multichannel biosignals have been studied in a number of works. However, these works were generally based on the MMV model. They only exploited common sparsity profiles among channel signals, but did not exploit the inter-channel correlation.  It is
shown \cite{Zhang2011IEEE} that if ignoring the correlation, most MMV-model based CS algorithms will have degraded recovery performance, especially in the presence of high inter-channel correlation. In the two EEG datasets used in our experiments, the inter-channel correlation between $\mathbf{Z}_{\cdot,i}$ and $\mathbf{Z}_{\cdot,j}$ is very high, generally above 0.9 (when $|i-j|\leq 5$). Thus it is not difficult to understand why ISL0 had poor
performance in the experiments. In fact, in the two experiments if STSBL-EM was performed without exploiting the inter-channel correlation (i.e. setting $\mathbf{B}=\mathbf{I}$), the BCI classification rate and the drowsiness estimation were very poor, even poorer than those by BSBL-BO.

%
%
%
%
%

Therefore, exploiting the inter-channel correlation is necessary in CS of multichannel signals; ignoring it can seriously deteriorate CS algorithms' performance. This also indicates the importance of our work in developing the STSBL-EM algorithm which can exploit the correlation.

\section{Conclusions}

We proposed a spatiotemporal sparse Bayesian learning algorithm for energy-efficient compressed sensing of multichannel signals. In contrast to existing compressed sensing algorithms, it not only exploits  correlation structures within a single channel signal, but also exploits inter-channel correlation. It  has much better recovery performance than state-of-the-art algorithms. Its speed is relatively stable even when the channel number significantly changes. Experiments on SSVEP-based BCI and EEG-based driver's drowsiness estimation showed that when using the proposed algorithm, the BCI classification rate and the drowsiness estimation on recovered signals were almost the same as those on original signals, even when the signals were compressed by 80\%.

Since the  algorithm takes root in Bayesian basis selection, it can be used in many other applications, such as feature selection, source localization, and sparse representation.

\bibliographystyle{IEEEtran}

\bibliography{bib_Zhilin}

\begin{biography}[{\includegraphics[width=1.0in,height=1.25in,clip,keepaspectratio]{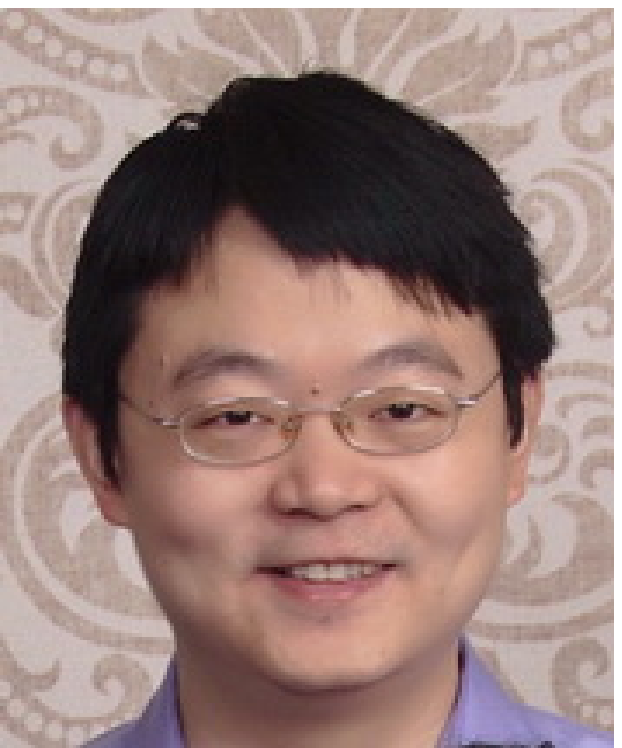}}]{Zhilin Zhang} received the Ph.D. degree in electrical engineering from University of California at San Diego in 2012. He is currently a senior research engineer in the Emerging Technology Lab in Samsung Research America -- Dallas.

His research interests include sparse Bayesian learning, sparse signal recovery, signal separation and decomposition, machine learning, and their applications to biomedicine, healthcare, and smart-home. He has authored or coauthored about 40 peer-reviewed journal and conference papers.

He is a technical committee member in Bio-Imaging and Signal Processing of the IEEE Signal Processing Society (from January 2014  to December 2016), and a technical program committee member of a number of international conferences.

He received Excellent Master Thesis Award in 2005, Second Prize in College Student Entrepreneur Competition (on fetal heart rate monitor) in 2005, and Samsung Achievement Award in 2013 and 2014.
\end{biography}

\begin{biography}[{\includegraphics[width=1.0in,height=1.25in,clip,keepaspectratio]{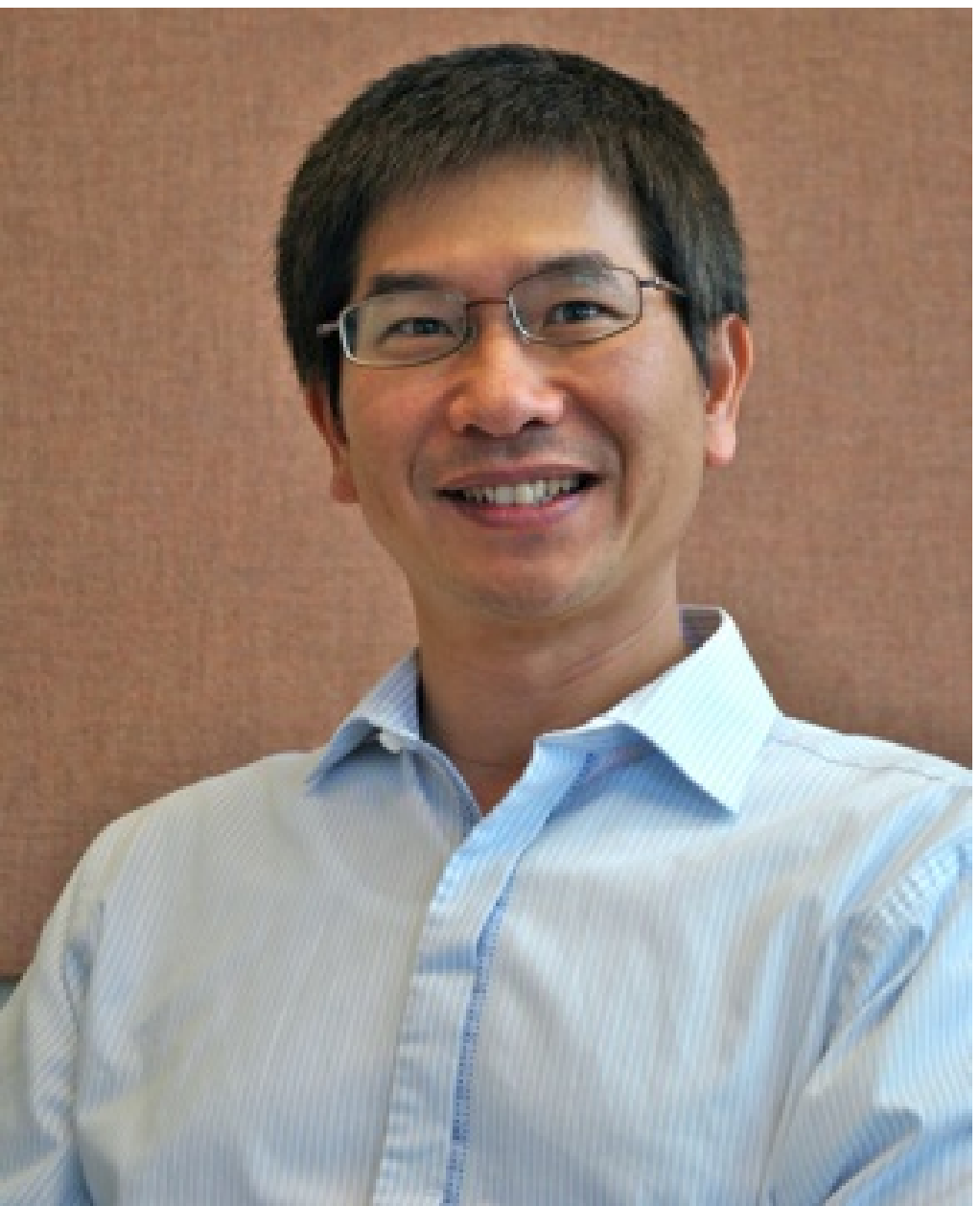}}]{Tzyy-Ping Jung} received
the B.S. degree in electronics engineering from
National Chiao-Tung University, Taiwan, in 1984,
and the M.S. and Ph.D. degrees in electrical
engineering from The Ohio State University,
Columbus, in 1989 and 1993, respectively.

He is currently a research
scientist and the co-director of the Center for
Advanced Neurological Engineering, Institute of Engineering in Medicine,
University of California at San Diego (UCSD). He is also an associate
director of the Swartz Center for Computational Neuroscience, Institute
for Neural Computation, and an adjunct professor of the Department of Bioengineering at
UCSD. In addition, he is a professor of Department of Computer
Science, National Chiao Tung University, Hsinchu, Taiwan.

His research
interests are in the areas of biomedical signal processing, cognitive
neuroscience, machine learning, time-frequency analysis of human EEG,
functional neuroimaging, and brain computer interfaces and interactions.

He received the Unsupervised Learning Pioneer Award from the Society for Photo-Optical Instrumentation Engineers in 2008. He is currently an Associate Editor of IEEE Transactions on Biomedical Circuits and Systems.

\end{biography}

\begin{biography}[{\includegraphics[width=1.0in,height=1.25in,clip,keepaspectratio]{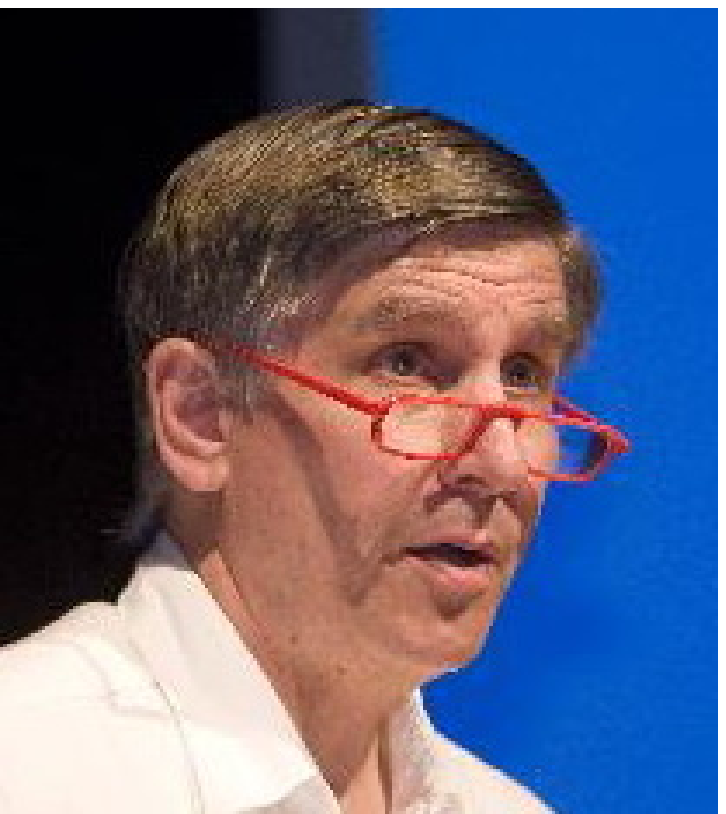}}]{Scott Makeig} received the B.S. degree, ``Self in Experience",
from the University of California Berkeley,
Berkeley, in 1972 and the Ph.D. degree in music
psychobiology from the University of California at
San Diego (UCSD), La Jolla, in 1985.

After spending a year in Ahmednagar, India, as a
American India Foundation Research Fellow, he
became a psychobiologist at UCSD, and then a
research psychologist at the Naval Health Research
Center, San Diego. In 1999, he became a staff scientist at the Salk
Institute, La Jolla, CA, and moved to UCSD as a research scientist in 2002 to
develop and direct the Swartz Center for Computational Neuroscience.

His
research interests are in high-density electrophysiological signal processing
and mobile brain/body imaging to learn more about how distributed
brain activity supports human experience and behavior.

\end{biography}

\begin{biography}[{\includegraphics[width=1.0in,height=1.25in,clip,keepaspectratio]{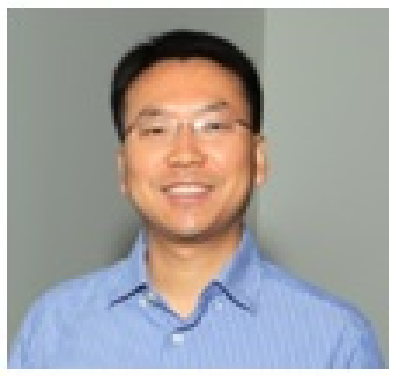}}]{Zhouyue Pi} is a Senior Director at Samsung Research America in Dallas, Texas, where he leads the Emerging Technology Lab doing research in next generation mobile devices, smart home solutions, and mobile health technologies.

Before joining Samsung in 2006, he was with Nokia Research Center in Dallas and San Diego, where he led 3G wireless standardization and modem development for 3GPP2 1xEV-DV, 1xEV-DO, and Ultra Mobile Broadband (UMB). In 2006 -- 2009, he was a leading contributor to Samsung's 4G standardization efforts in 3GPP LTE and IEEE 802.16m, and to IEEE 802.11ad for 60 GHz communication. In 2009 -- 2012, he pioneered 5G mm-wave massive MIMO technology and led the development of the world's first baseband and RF system that demonstrated the feasibility of Gbps mobile communication at 28 GHz.

He has authored more than 30 technical journal and conference papers and is the inventor of more than 150 patents and applications. He holds a B.E. degree from Tsinghua University (with honor), a M.S. degree from the Ohio State University, and an MBA degree from Cornell University (with distinction). He is a Senior Member of IEEE.
\end{biography}

\begin{biography}[{\includegraphics[width=1.0in,height=1.25in,clip,keepaspectratio]{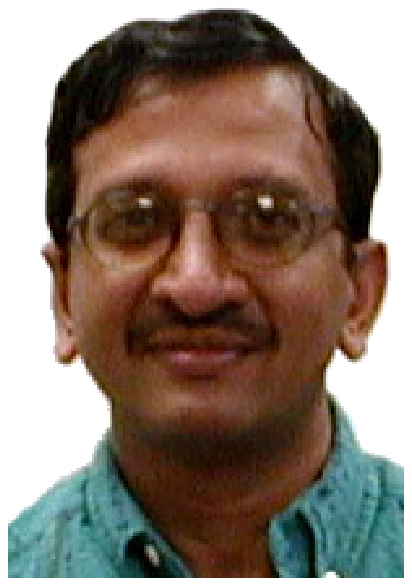}}]{Bhaskar D. Rao} received the Ph.D. degree
from the University of Southern California in
1983. Since 1983, he has been with the University of
California at San Diego (UCSD),
where he is currently a professor with the Department of Electrical and Computer
Engineering and the holder of the Ericsson endowed chair in
wireless access networks.

His interests are in the areas of digital
signal processing, estimation theory, and optimization theory, with
applications to digital communications, speech signal processing, and
human-computer interactions. He
has been a member of the Statistical Signal and Array Processing
Technical Committee, the Signal Processing Theory and Methods
Technical Committee, the Communications Technical Committee of the
IEEE Signal Processing Society.

His work has received several paper awards.
His paper received the Best Paper Award at the 2000 Speech Coding
Workshop and his students have received the Best Student Paper Awards at both
the 2005 and 2006 International Conference on Acoustics, Speech and
Signal Processing (ICASSP), as well as the Best Student Paper Award
at Neural Information Processing Systems Conference (NIPS) in 2006. A paper he co-authored  with B. Song and  R. Cruz
received the 2008 Stephen O. Rice Prize Paper Award in the Field of
Communications Systems. A paper co-authored by him and his student received the 2012 Signal Processing Society (SPS) Best Paper Award.
He was elected to the IEEE Fellow in 2000
for his contributions to the statistical analysis of subspace algorithms for harmonic retrieval.
\end{biography}

\end{document}